\documentclass[acmsmall,screen=True,anonymous=False,review=False]{acmart}

\AtBeginDocument{%
  \providecommand\BibTeX{{%
    \normalfont B\kern-0.5em{\scshape i\kern-0.25em b}\kern-0.8em\TeX}}}

\usepackage[utf8x]{inputenc}
\usepackage{natbib}
\def\citepos#1{\citeauthor{#1}'s (\citeyear{#1})}

\begin{document}

\title[Governance Capture in a Self-Governing Community]{Governance Capture in a Self-Governing Community: A Qualitative Comparison of Serbo-Croatian Wikipedias}

\author{Zarine Kharazian}
\email{zkharaz@uw.edu}
\orcid{0000-0003-1937-7804}

\author{Kate Starbird}
\email{kstarbi@uw.edu}
\orcid{0000-0003-1661-4608}

\author{Benjamin Mako Hill}
\email{makohill@uw.edu}
\orcid{0000-0001-8588-7429}
%\email{webmaster@marysville-ohio.com}
\affiliation{%
  \institution{University of Washington}
%  \streetaddress{P.O. Box 1212}
  \city{Seattle}
%  \state{Washington}
  \country{USA}
  \postcode{98195}
}

\renewcommand{\shortauthors}{}

\begin{abstract}
What types of governance arrangements makes some self-governed online groups more vulnerable to disinformation campaigns? To answer this question, we present a qualitative comparative analysis of the Croatian and Serbian Wikipedia editions. We do so because between at least 2011 and 2020, the Croatian language version of Wikipedia was taken over by a small group of administrators who introduced far-right bias and outright disinformation; dissenting editorial voices were reverted, banned, and blocked. Although Serbian Wikipedia is roughly similar in size and age, shares many linguistic and cultural features, and faced similar threats, it seems to have largely avoided this fate. Based on a grounded theory analysis of interviews with members of both communities and others in cross-functional platform-level roles, we propose that the convergence of three features---high perceived value as a target, limited early bureaucratic openness, and a preference for personalistic, informal forms of organization over formal ones---produced a window of opportunity for governance capture on Croatian Wikipedia. Our findings illustrate that online community governing infrastructures can play a crucial role in systematic disinformation campaigns and other influence operations.
\end{abstract}




\received{January 2023}
\received[revised]{July 2023}
\received[accepted]{October 2023}

 \maketitle

% \fontsize{12pt}{24pt}
% \selectfont

\section{Introduction}

Self-governed online groups like Wikipedia play a critical role in the contemporary information environment and are often seen as bulwarks against misinformation and disinformation. For example, visitors to English YouTube videos arguing that the earth is flat are automatically linked to the ``Flat Earth'' article on English Wikipedia that explains that ``Flat Earth is an archaic and scientifically disproven conception of the Earth's shape as a plane or disk.''\footnote{\url{https://en.wikipedia.org/wiki/Flat_Earth} (\url{https://perma.cc/P2RR-ANXM})} Given Wikipedia's open editing policy, this arrangement is remarkable. Unlike centrally controlled commercial platforms like YouTube itself, self-governed groups like Wikipedia have participatory governance structures that make it possible for governance to be captured by outside groups promoting coordinated disinformation campaigns with very few resources. While some communities, like English Wikipedia, appear to be remarkably resistant to these types of attacks, this resistance is not universal \cite{mcdowell.vetter_2020}. Why do some self-governing communities succeed in developing institutions that are resistant to these types of campaigns while others revert to forms of rule with poor information integrity outcomes?

A recent case involving the Croatian language edition of Wikipedia offers an opportunity to explore this question. The facts of this case have been well-documented in several sources, including a series of public discussions by Wikipedia editors,\footnote{\url{https://meta.wikimedia.org/wiki/Requests_for_comment/Hard_line_nationalism_on_the_Croatian_Wikipedia} (\url{https://perma.cc/EV63-ZJ8R}), \url{https://meta.wikimedia.org/wiki/Requests_for_comment/Administrator_abuse_on_the_Croatian_Wikipedia} (\url{https://perma.cc/LTP5-M6PU}), \url{https://meta.wikimedia.org/wiki/Requests_for_comment/Site-wide_administrator_abuse_and_WP:PILLARS_violations_on_the_Croatian_Wikipedia} (\url{https://perma.cc/RR39-GLNP})} reporting in regional media \cite{milekic_2018, jergovic_2018, caucaso_2018}, and Wikipedia's own volunteer-run newsletter \cite{gregorb_2023}. In summary, over the span of a decade, a small group of editors seized control of Croatian Wikipedia and systematically introduced content that reflected narratives commonly associated with the Croatian far-right. In doing so, they violated a series of Wikipedia's fundamental principles related to neutrality, verifiability, and reliability.

In 2021, the Wikimedia Foundation (WMF) published a retrospective ``disinformation assessment'' on the state of Croatian Wikipedia that
``revealed numerous examples of systemic, deeply-rooted bias and disinformation'' on Croatian Wikipedia and characterized the situation as an instance of ``project capture'' \cite{wikimediafoundationtrust&safetydisinformationteam_2021}.
In part, the report attributed Croatian Wikipedia's capture to a unique situation in which there were distinct Wikipedia editions for the standardized national variants of a pluricentric language: Bosnian-Croatian-Montenegrin-Serbian (BCMS), sometimes referred to as Serbo-Croatian \cite{vrabec2021bosnian}. This explanation, however, raises the question of why Serbian and Bosnian Wikipedia did not appear to suffer Croatian's fate.

We present a qualitative analysis of interview data with a range of participants in Croatian and Serbian Wikipedia and in the broader Wikipedia community. In doing so, our research seeks to answer \textit{why} and \textit{how} Croatian Wikipedia succumbed to capture when similar projects did not. This work has important implications for the study of coordinated disinformation campaigns and the governance of social computing systems in general. We contribute to the disinformation literature in several ways. While most disinformation research has focused exclusively on features of content, audiences, and on processes of dissemination \cite{wilson.etal_2018, prochaska.etal_2023, lukito_2020}, we take an organizational and institutional approach. In contrast to studies that have focused on failures to maintain information integrity \cite{albadi.etal_2019, velasquez.etal_2021}, our comparative approach means that our study provides rare insight into processes of resilience. Our work is also unusual in that it focuses on Wikipedia rather than social media platforms like Twitter \cite{zannettou.etal_2019} and Facebook \cite{bailey.etal_2021}. Our research also contributes to the rapidly growing literature on self-governance in social computing systems by providing one of the first comparative qualitative analyses of institutional development from birth to maturity. It also connects this body of research on self-governance to an important social outcome (i.e., disinformation). Finally, our work contributes to both disinformation and social computing scholarship more broadly by providing insight into rarely studied non-English contexts.

\section{Related Work}
\label{sec:related-work}
Our work builds upon three areas of ongoing research. The first involves work on community governance, moderation, and information quality in social computing systems. The second concerns how participatory disinformation campaigns manifest on online platforms, including peer production projects like Wikipedia. The third is the significant body of literature in political economy on the phenomenon of governance capture and its applications to the study of online community governance.

\subsection{Governance and Information Quality}

Drawing from social psychology and behavioral economics, \citepos{kraut.resnick_2012} book \textit{Building Successful Online Communities} describes the regulation of behavior as one of the most important goals of every online community and a central organizing theme in social computing scholarship. Their influential approach focuses on a top-down design-based approach to encouraging good behavior and discouraging bad behavior on the part of community members by designing incentives and sanctions through reputation systems, explicit rules, mechanisms to verify user identities, and so on. 

A contrasting approach advocated for by \citet{frey.etal_2019} draws from the work of Nobel laureate Elinor Ostrom to emphasize the bottom-up empirical study of institutions and governance systems. \citeauthor{frey.etal_2019} argue that because \citeauthor{kraut.resnick_2012} assume that the goal of social computing scholarship is to inform a ``designer'' so that they might effectively shape a community according to their desires, Kraut and Resnick imply a (highly centralized) form of governance. \citet{frey.etal_2019} cite \citet{ostrom_1990} to argue that important differences in governance arrangements are precluded by this approach. They argue that these differences can deeply shape outcomes and experiences in complex social computing systems like organizational resilience.

A growing number of studies in CSCW and social computing have focused on the the development of governance structures \citep{seering_2020} and described the emergence of community rules and norms to govern behavior \citep{fiesler.etal_2018, omahony.ferraro_2007, forte.etal_2009}. Some work has focused specifically on the experiences of community members in responding to and defending against coordinated attacks and networked harassment \cite{marwick_2021, herring.etal_2002, massanari_2017, han.etal_2023}. 

One of the most popular types of settings to study the emergence of these governance structures is online peer production. Peer production is a term coined by Yochai Benkler to describe a model of producing knowledge goods through the mass aggregation of many small contributions from diversely motivated individuals \citep{benkler_2002, benkler_2006, benkler.etal_2015}. Although peer production can happen in a range of commercial and non-commercial platforms, researchers have focused disproportionately on Wikipedia {\cite{viegas.etal_2007, noack.etal_2010, jemielniak_2015}, the world's most successful example of the organizational form \citep{benkler.etal_2015}. In peer production, the community's ability to regulate behavior directly affects the quality of the information goods produced. For example, in a study on the r/AskHistorians subreddit, \citet{gilbert_2020} considers how the subreddit's moderation practices fosters trust in the information shared on the subreddit. In a recent book, \citet{bruckman_2022} focuses on the process of social knowledge construction on Wikipedia and explain that Wikipedia's institutional arrangements shape information quality and reliability.

\subsection{Disinformation and Peer Production}
\label{disinformation-and-peer-production}

\textit{Disinformation}---the intentional spreading of false or misleading information---is a tactic that actors utilize to alter and distort the information environment \citep{bittman_1981}. 
More recently, many scholars have shifted to the term \textit{influence operations} to describe campaigns that in fact blend elements of truth, rationalization, speculation, and propaganda \cite{beers.etal_2022}. Contemporary research on disinformation campaigns has focused primarily on their online elements and how they manifest on commercial social media platforms such as Facebook, Twitter \citep{zhang.etal_2021}, and to a lesser extent, YouTube \citep{hussein.etal_2020, hussain.etal_2018}. The affordances of these platforms---personalized recommendation systems and engagement-based interaction mechanisms in particular---make them particularly vulnerable to influence efforts that can harness the dynamics of virality to promote certain messaging to specific audiences at scale. Additionally, popular conceptions of disinformation campaigns tend to focus on highly coordinated top-down efforts such as Russia's Internet Research Agency's efforts to influence political discourse in 2016 \cite{boatwright.etal_2018, diresta.etal_2019, lukito_2020}. CSCW researchers, however, have previously highlighted the participatory nature of these campaigns, which often take shape as collaborations between government agents, political activists, and everyday people \cite{starbird.etal_2019}. 

Wikipedia occupies a unique role in the online disinformation ecosystem. The platform as a whole has proved to be relatively robust to the types of virality-fueled disinformation campaigns that have plagued commercial social platforms. This robustness is in part due to the fact that Wikipedia is run by a nonprofit organization and does not rely on engagement mechanisms that underlie the for-profit, advertiser-driven model, as well as its culture of collaboration and transparency. Wikipedia's reputation as a source of generally timely and authoritative information on a range of topics has in turn made it a critical tool in the arsenal of fact-checking initiatives implemented by the commercial tech giants Facebook and YouTube \citep{saez-trumper_2019, matsakis_2018}. It has also led some commentators to hail Wikipedia's approach to community governance and content moderation as a viable alternative model to that of commercial social media \citep{cooke_2020}.

But as one of the top ten most-visited websites in the world, a massive repository of human knowledge, and a resource for fact-checking initiatives, Wikipedia also remains an attractive target for influence operations \cite{aragon.saez-trumper_2021}. Previous research has documented efforts to degrade the integrity of Wikipedia's content through vandalism \citep{tramullas.etal_2016, geiger.ribes_2010}, hoaxes \citep{kumar.etal_2016}, sockpuppets \citep{solorio.etal_2014}, and paid promotion \citep{joshi.etal_2020}. Other research has focused on problematic behaviors that negatively affect the community of contributors such as trolling and harassment \citep{shachaf.hara_2010}. While much of this work has engaged in close reading of article edit histories and inter-editor interactions, we know of no work that has considered how distributed influence operations \citep{starbird.etal_2019} target, become deeply engaged with, and are facilitated by institutional and organizational arrangements within peer production communities like Wikipedia.

\subsection{Capture of Self-Governing Institutions}

In the realm of traditional institutions, a significant body of work in political economy looks at the phenomenon of state capture: the process by which an interest group, firm, or some other minority constituency co-opts the governance mechanisms of a state in service of its personal, ideological, or financial interests \cite{innes_2014, tudoroiu_2015, perry.keil_2018, kopecek_2016}. In the seminal paper on the phenomenon, \citet{hellman.etal_2003} argue that opportunities for state capture by private elites abounded in transition economies, such as the newly independent states of the former Soviet Union, where liberal economic development significantly outpaced the development of institutional reforms focused on building a requisite regulatory capacity. 

While peer production projects like Wikipedia do not possess the centralized bureaucratic structures of nation states, they still employ forms of bureaucratic organization, including formal and informal policies \cite{butler.etal_2008, joyce.etal_2013} and leadership hierarchies \cite{jemielniak_2015, zhu.etal_2012a}, as a means to organize and sustain collective action. These types of communities are inherently vulnerable: although many may aspire to democratic ideals, the underlying technical systems on which they rest frequently result in the opposite. In the context of online peer production, \citet{shaw.hill_2014} find that most projects become oligarchic over time as a small group of initial leaders consolidate power. Similarly, \citet{schneider_2022} argues that the technical affordances of the technological systems on which most online communities rely encourage the emergence of an ``implicit feudalism'' where power is concentrated in the hands of a few users with greater access privileges, such as administrators and moderators, and suggests that this represents the dominant governance ideology across digital platforms. Given these forces, online communities that sustain democratic practices are an exception rather than the default. 

\section{Empirical Setting: BCMS Wikipedias}
\label{empirical-setting}

\begin{table}
% \resizebox{\columnwidth}{!}{%
\caption{Details about four different BCMS Wikipedia editions.}
\label{tab:wikiinfo}
\begin{tabular}{@{}lllll@{}}
\toprule
                             & Serbo-Croatian & Serbian  & Croatian & Bosnian  \\ \midrule
Launch Date                  & Jan 2002       & Feb 2003 & Feb 2003 & Dec 2002 \\
Number of Articles           & 457K           & 662K     & 213K     & 90K      \\
Active Editors (monthly avg since 2016) & 40             & 318      & 181      & 44       \\ 
Unique Devices - readers (monthly avg since 2016)  & 2M             & 4M      & 4M      & 1M       \\ 
\bottomrule
\end{tabular}%
% }
\end{table}

Our empirical setting is the Wikipedia editions in Croatian, Serbian, Bosnian, and what is referred to as the ``Serbo-Croatian'' edition. While the first three were created in 2002--2003 to focus on national variants of BCMS, the fourth predated the others and seeks to serve all BCMS language speakers.\footnote{There is currently no Wikipedia edition for Montenegrin, despite longstanding proposals to create one. A test version of Montenegrin Wikipedia has existed since December 2017 on Wikimedia Incubator. \url{https://incubator.wikimedia.org/wiki/Wp/cnr} (\url{https://perma.cc/AE23-DLZZ})} 
Together, these projects represent four of over 300 Wikipedia language editions that vary in terms of size, community capacity, and sociopolitical context \citep{khatri.etal_2022}. 
Details on all four projects are shown in Table \ref{tab:wikiinfo}.

Wikipedia's governance model is what \citet{ostrom_1990} describes as ``polycentric.'' There are several fundamental principles that are shared across all projects. Foremost among these are the ``Five Pillars.''\footnote{\url{https://en.wikipedia.org/wiki/Wikipedia:Five_pillars} (\url{https://perma.cc/W92J-8Y3R})} There are also a small number of individuals and organizations with technical and administrative responsibility across different language editions. Compared to social media platforms, this centralized governance layer is incredibly thin. Individual language communities enjoy a large degree of autonomy in governance, including the ability to write, interpret, and enforce their own policies, run their own formal non-profit organizations, and choose their own leaders \cite{hwang.shaw_2022,forte.etal_2009, butler.etal_2008, steinsson_2023}. 

Existing comparative analysis of Wikipedia language editions beyond the English language edition has focused on how Wikipedia's collaborative model plays out in non-Western contexts \cite{bipat.etal_2021, hickman.etal_2021, khatri.etal_2022}. One common thread across these studies is the socially and culturally specific ways in which in which Wikipedia's Five Pillars are negotiated differently on particular language editions.

Several studies have examined this in the context of the BCMS environment. \citet{mujadzevic_2021} describes the post-Yugoslav online environment as ``one of the very few most explosive online memorial landscapes in the post-socialist space,'' and points to online encyclopedias driven by ``small groups of dedicated users focusing on onesided and politically charged interpretations of painful and traumatic historic events in the 20th century [that are] of service to the nationalist identity mobilisation.'' \citet{bilic.bulian_2014} compared articles related to the Republic of Kosovo on the Serbian, Croatian, and English Wikipedias and found frequent conflicts between editors on the former two who privileged an encyclopedic identity and those that privilege a nationalist identity. In another comparative study of Wikipedia articles about the 1995 Srebrenica massacre of Bosniaks by the Bosnian Serb army, \citet{rogers_2013} found that the English, Dutch, Serbian, Croatian, and Serbo-Croatian Wikipedia editions each presented a distinct account of the same series of historical events, with the Dutch article focusing on the ``fall of Srebenica,'' the English version referring to the ``massacre'' that occurred afterward, the Croatian and Bosnian versions describing the events as genocide, and an early version of the Serbian version downplaying the massacres before devolving into ``edit warring.''

While Wikipedia language editions have, to a certain extent, struggled to balance cultural specificity with encyclopedic principles, often resulting in contentious editing disputes, there is near consensus within the broader Wikipedia communities that the extent of historical revisionism on the Croatian Wikipedia from 2011-2020 violated the Five Pillars in an otherwise unprecedented way. One example was the article on the Jasenovac concentration camp, the third largest World War II concentration camp in Europe which was run by the Ustase regime \cite{mccormick_2014}. The Ustase was a Croatian fascist movement that was the ruling regime of the Independent State of Croatia (1941-1945), a Nazi puppet state that controlled Axis-occupied parts of Yugoslavia during World War II \cite{korb2010nation}. The Ustase regime was responsible for massacring hundreds of thousands of Serbs, Jews, Roma, and political dissidents \citep{mujadzevic_2021}. Today, the regime is glorified by the modern Croatian far-right \citep{pavlakovic_2008}. While Wikipedia editions in English, German, French, and Spanish all accurately referred to Jasenovac as a ``concentration and extermination camp,'' the Croatian Wikipedia versions of the article published between 2017 to 2020 referred to it as a ``collection and labor camp'' and engaged in numerous false equivalencies in how it presented sources. For example, it gave equal weight to professional academic and historical accounts of the number of victims at Jasenovac and the much lower estimates offered by known Holocaust and Jasenovac denialists \cite{wikimediafoundationtrust&safetydisinformationteam_2021}. In 2018, the regional news outlet \textit{Balkan Insight} published an article titled, ``How Croatian Wikipedia Made a Concentration Camp Disappear,'' describing large portions of Croatian Wikipedia as downplaying the crimes and number of victims of the Ustase regime \cite{milekic_2018}. The article also cited several Croatian historians who did not consider Croatian Wikipedia to be a reliable source on account of its far-right nationalist bias.

By 2020, owing largely to efforts by the volunteer editor community, many of the administrators of Croatian Wikipedia had been removed from their positions of power and, in the most egregious cases, banned from Wikipedia. In 2021, the WMF published a retrospective ``disinformation assessment'' on the state of Croatian Wikipedia. The report noted that a small group of administrators and users with other elevated rights ``intentionally deflected legitimate concerns about content bias and/or problematic behavior'' through the use of ``well-known disinformation tactics, including relativisation of facts, whataboutism, discreditation of other participants, and outright bullying'' \citep{wikimediafoundationtrust&safetydisinformationteam_2021}. 

The report suggested that Croatian Wikipedia had been ``captured'':

\begin{quote}
Project capture of Croatian language Wikipedia has exposed---and exploited---a weakness in Wikipedia's model of community self-governance. Echoing the phenomenon of state capture, one of the most pressing issues of today's worldwide democratic backsliding, the case of [Croatian Wikipedia] has demonstrated that both community and content can and will decline if institutions are taken over by an organised and ideologically aligned group \citep{wikimediafoundationtrust&safetydisinformationteam_2021}.
\end{quote}

\noindent In part, the report traced the roots of Croatian Wikipedia's capture to the decision to created separate ``national'' Wikipedia editions corresponding to the BCMS language variants. The creation of separate editions corresponding to national variants of a language appears to be unique to the BCMS environment within Wikipedia. The 2003 split produced a series of national Wikipedia projects in which ethnonationalism had greater salience and resonance, and that were ultimately conducive to capture by an ethnonationalist contingent of contributors.

However, this explanation is incomplete at best. While historical revisionism and nationalist bias are present on all four projects to varying degrees \cite{rogers_2013}, Croatian Wikipedia alone experienced a complete takeover of its mechanisms for decision-making, dispute resolution, and member accountability \cite{wikimediafoundationtrust&safetydisinformationteam_2021}. Why did the Serbian, Bosnian, and Serbo-Croatian Wikipedia not appear to suffer Croatian Wikipedia's fate? Our analysis suggests a more full answer relies on a systematic understanding of differences in governance across the projects.

\section{Methods}
\label{methods}

There are many ways to contribute to Wikipedia. Some Wikipedians contribute to cross-project governance, others engage in cross-wiki monitoring efforts, and others operate at the level of individual projects. Many contributors may also serve roles at several of these levels, such as a user with global rollbacker privileges who also serves as an administrator for a Wikipedia language edition. As a result, different users have different perspectives on the issue of project capture. We recruited both participants who worked primarily at the global or cross-wiki level and those whose contributions were mostly focused on a single project: either Croatian or Serbian Wikipedia. 

The first group of interviewees were recruited from the following groups: (1) \textit{stewards}: elected volunteers who are appointed across Wikipedia editions with elevated global privileges; %  and are the frontline defense for many conduct enforcement issues. 
(2) \textit{WMF staff}: employees of the WMF, often focused on trust \& safety and disinformation issues, who rarely intervene in matters of local governance on Wikipedia projects;
(3)\textit{members of cross-wiki monitoring user groups}:  self-organized volunteer groups engaging in some form of cross-wiki monitoring or patrolling of vandalism across language editions (e.g., the Small Wiki Monitoring Team, the Countervandalism Network).
Additionally, we also interviewed one contributor who was an editor and rollbacker in English Wikipedia but had also previously contributed to several other editions, including Dutch, German, and Spanish.

The second group of interviewees were Wikipedia contributors who were involved directly in the Serbian or Croatian Wikipedias as editors, administrators, or other users with elevated privileges. While this group of users did not typically have the breadth of experience across many different projects, they had in-depth knowledge of the language edition or editions to which they regularly contributed. These participants spoke about their knowledge and experiences on the four BCMS Wikipedias. Although individual engagement would ebb and flow over time, every one of these interviewees had been involved during the decade that Croatian Wikipedia had been captured and several had been involved in the period before as well.

In total, we conducted 15 interviews, 14 of which were conducted in English over video conferencing software by the lead author between April 2022 and October 2022. One interview was conducted by email, at the interviewee's request. The audio interviews lasted from 25 to 116 minutes, with an average interview length of 67 minutes.
All audio interviews were recorded and transcribed by a student research team led by the first author. The interview questions were open-ended, but designed to elicit observations about interviewees' experiences with and perceptions of local and global Wikipedia governance systems, community capacity, and resilience to project capture threats. When relevant, the protocol included follow up questions to elicit specific examples from the interviewee's personal experiences. A copy of the interview protocol is included in Appendix \ref{sec:protocol}.

To recruit interviewees, we employed a combination of statistically non-representative stratified sampling \citep{trost_1986} and snowball sampling. Initial participants were recruited from the Croatian and Serbian Wikipedia language editions by surveying the list of former and current administrators in each project as well as editors that have been active in talk page discussions relevant to community governance issues. Participants were also recruited from public lists of active stewards and those listed as involved with various cross-wiki anti-vandalism efforts, such as the Small Wiki Monitoring Team. Additionally, a WMF staff member with knowledge of disinformation issues was also recruited. Finally, we used snowball sampling to recruit individuals who may not have been as publicly active in discussions but who were still knowledgeable about the topic.

% This sampling strategy does not result in a statistically representative sample: we deliberately selected interviewees with more interest in and knowledge of community governance issues on Wikipedia than the average editor, as evidenced by either their formal roles or ongoing public participation in such discussions. Ultimately, however, this resulted in a group of informants who had a certain level of domain expertise, having thought about community governance issues in depth, often for years or even over a decade, before our discussions with them.

Given the sensitive nature of the topic, the relatively small numbers of active contributors in each of these categories, and the specificity of some participants' experiences, we took additional steps to prevent the deanonymization of our interview data. Rather than reporting detailed demographic information for each Participant ID, we have opted to report aggregate information of the roles represented across the interview pool and the primary language editions to which interviewees reported being active contributors. These details are shown in Table \ref{tab:participants}.

\begin{table}[t]
%\resizebox{\columnwidth}{!}{%
\caption{List of interview participants in four ``buckets'' that describe the background, skill, and wiki-level experiences of each subject. For editors of the Croatian and Serbian language editions, approximate years of engagement with the project are included in parentheses. Groups are aggregated in this way to preserve the anonymity of research subjects.}
\label{tab:participants}
\begin{tabular}{p{0.38\textwidth}p{0.32\textwidth}p{0.23\textwidth}}
\toprule
  \textbf{Represented Roles} &
  \textbf{Primary Language Editions} &
  \textbf{Participant IDs} \\
  \midrule
\textbf{Global \& Cross-wiki} &&\\
Steward, WMF staff, Small Wiki Monitoring Team members; other cross-wiki efforts & English, Simple English, Dutch, Czech 
&
  P01, P02, P04, P11, P12 \\
\textbf{Local} && \\
Editor, admin      & Croatian & P05 (2020-) \\ 
&  & P06 (2005-) \\ 
&  & P09 (2016, 2020) \\
&  & P13 (2013-) \\
&  & P14 (2004-) \\
Editor, admin      & Serbian  & P07 (2010-) \\ 
&  & P08 (2005-) \\
&  & P10 (2004-) \\ 
&  & P15 (2018-) \\
Editor, rollbacker & English  & P03                     \\ \bottomrule
\end{tabular}%
\end{table}

Our analysis proceeded following \citepos{charmaz_2014} approach to grounded theory. We engaged in a four-step process. First, a student research team led by the first and last authors conducted line-by-line open coding of the transcripts using the open-source qualitative data software Taguette. Codes were discussed and compared across interviews during weekly meetings. In the second step, the lead author conducted focused coding, identifying the most useful initial codes and combining them to to synthesize data across incidents. The third step was axial coding: using the concept visualization software Miro, the lead author collapsed the focused codes from interviews into categories and arrived at themes that surfaced across interviews. In the final step, the lead author synthesized these themes into a conceptual model, presented as Figure \ref{fig:capture}. Consistent with a grounded theory approach, the data collection and the four steps of analysis were conducted in an iterative fashion, moving back and forth between data collection and the different stages of analysis as the research team refined the theoretical categories that emerged from the data through memoing.

\section{Findings}
\label{findings}

Our analysis of the interviews yielded three factors that contributed to Croatian Wikipedia's capture. We present these findings as three propositions that, together, form an explanation for why and how Croatian Wikipedia descended in to capture, while other BCMS editions, despite similar initial conditions, did not.

\subsection{Proposition 1: Perceived Value as a Target}
\label{proposition-1-percieved value} 

Online communities can be valuable in many ways. For example, the communities that sustain Wikipedia projects provide public value by producing an information good from which society benefits \cite{meijer.boon_2021} as well as social and psychological value to their contributors \cite{rafaeli.ariel_2008}. Our interviewees suggested that some types of value contributed to projects' strategic use as targets and increased the risk of attempted capture. 

Although our interviewees' experience was largely with the Serbian and Croatian Wikipedias, all of our interviewees from either project had some familiarity with the dynamics on the other two adjacent language projects: Serbo-Croatian and Bosnian. Some participants even edited more than one of these projects. Through discussions with these participants about the broader BCMS language environment on Wikipedia, it became apparent that the four projects were not equally attractive targets for a nationalist disinformation campaign. In this respect, Croatian and Serbian fulfilled two criteria that Serbo-Croatian and Bosnian did not: they had \textit{both} a critical mass in terms of potential audience as well as a community in which national narratives resonated. Together, these two qualities increased the Croatian and Serbian Wikipedias' perceived strategic value, making them targets for actors looking to promote revisionist constructions of national pasts.

\subsubsection{Audience Size.}

Participants consistently noted that Serbian and Croatian had a larger readership than the Bosnian and Serbo-Croatian Wikipedias. While participants assessed the audience size on each language edition based on their own knowledge, their assessments were triangulated and confirmed with the quantitative measures of audience size made available via Wikistats and displayed in Table \ref{tab:wikiinfo}. In the case of Bosnian Wikipedia, the project's smaller audience size could likely be attributed to the significantly fewer number of Bosnian-language speakers (at roughly 2 million) \cite{worlddata.info_2023b}, forming, as P08 put it, ``the smallest natural community'' of the four projects.

Participants provided a different explanation, however, for why Serbo-Croatian, the first of the four projects, remained small in readership.
P13, who primarily edits Croatian Wikipedia but occasionally contributes to the Serbo-Croatian project, noted that Serbo-Croatian Wikipedia had the lowest readership among the four language editions, going as far as to call the project ``invisible'' to most Croatian internet users. P13 elaborated that unless ``you're really into Wikipedia,'' the average user who stumbles across Wikipedia results from a Google Search may not even know the Serbo-Croatian project exists, because the national Wikipedia editions are prioritized in search results based on location. P07, an editor who contributes to all four of the BCMS projects, similarly explained that ``nobody would look for the Serbo-Croatian Wikipedia... nobody calls the language [they speak] Serbo-Croatian.''

Prior work has shown that audience size incentivizes contributions to online communities producing information goods \cite{zhang.zhu_2011, teblunthuis.etal_2022, kraut.resnick_2012}. Audience size confers certain social benefits, primarily because it provides a greater opportunity for attention. For contributors to an information good like Wikipedia, the sense that one's contributions are finding an interested audience provides social satisfaction. While audience size has mainly been identified as a motivating factor for \textit{good-faith contributions}, we argue that it also plays a role in motivating \textit{damaging} behavior and bad-faith contributions. The potential to reach a large audience is valuable for motivated actors seeking to promote particular messaging. \citet{massanari_2017}, for example, documents how misogynistic activists exploited Reddit's ``karma'' points system to propel anti-feminist content to popularity across the site, indicating a deliberate attempt to promote their content to the largest possible audience. 

\subsubsection{National Resonance.} 

Our interviewees noted that Serbo-Croatian Wikipedia attracted a more niche audience than the other three projects. An editor from Serbian Wikipedia, P08, described it as a project for ``outcasts'' that did not fit in on any of the other three language editions that corresponded with a national identity:

\begin{quote}
On Serbo-Croatian Wikipedia, I just feel like there's no community. There's several disparate editors that edit on their own. It always sort of felt like a place for outcasts. If you don't belong in any of the three mainstream Wikipedias, then maybe this is for you. Maybe you're nostalgic, maybe you were banished from one of the others... to me, it felt like a compromise, but nothing else.
\end{quote}

\noindent P13 echoed this observation, noting that most people prefer to ``focus on their own national projects.'' They explained that Serbo-Croatian tended to serve as a project for ``refugees'' that tired of the nationalist agendas and ensuing neutrality disputes that plagued the Croatian, Serbian, and Bosnian Wikipedias. All of the interviewees familiar with the Serbo-Croatian projects classified Serbo-Croatian as, by far, the least contentious environment of the four language editions.

Disinformation campaigns often seek to exploit divisions on the basis of identity, whether those identities be ideological, social, racial, or ethnopolitical \cite{stewart.etal_2018}. Of the four Wikipedias in the BCMS language environment, editors on the Serbo-Croatian Wikipedia were the most likely to privilege an encyclopedic identity over a national one in the process of knowledge construction, an observation from our interviews that has also been corroborated by \citet{rogers_2013}. This ultimately made the project a considerably less attractive target than the Croatian and Serbian Wikipedias where nationalist narratives had the potential to resonate further.

\subsection{Proposition 2: Bureaucratic Openness}
\label{proposition-2-bureaucratic-openess}

As outlined in §\ref{proposition-1-percieved value}, the Serbian and Croatian Wikipedias shared characteristics that rendered them similarly valuable as targets for capture. The first difference in the institutional and organizational dimensions of the projects that appear to have bolstered the resilience of the former, while leaving the latter more vulnerable to project capture attempts, is \textit{bureaucratic openness}. A high degree of bureaucratic openness in Serbian Wikipedia early on appears to have played a critical role in increasing community capacity as well as diversity, two resources that helped the project deal with disinformation threats. In contrast, Croatian Wikipedia appeared not to prioritize openness early on in the project, and instead developed an insular bureaucratic culture characterized by the rise and entrenchment of user ``cabals'' that ruled over the project without wider community input or oversight. These differences in bureaucratic culture of the two Wikipedias were noted by all interviewees who had experience with the BCMS Wikipedia environment.

\subsubsection{Early Efforts to Expand Leadership Team on Serbian Wikipedia.}
\label{early-efforts}

Croatian and Serbian Wikipedia's trajectories may have diverged early on, as a result of contrasting priorities among both projects' early leaders. P10, a founder of Serbian Wikipedia, described Serbian Wikipedia's initial admin team as a group of educated professionals---many of them in academia---dedicated to the goal of creating an encyclopedia. But importantly, this group did not remain insular. 

On Wikipedia, ``bureaucrat'' status describes the ability to add or remove administrators. P10 describes how, early on, Serbian Wikipedia's initial team set up an open process for contributors to ascend to these elevated rights positions: ``My initial approach was, whoever stayed on Serbian Wikipedia and edited a month would get admin permissions, and whoever stayed for two months would get bureaucrat permissions.'' This strategy proved instrumental in attracting and retaining active newcomers to the Serbian Wikipedia, particularly to bureaucrat positions.

On Croatian Wikipedia, P10 asserts that ``something different happened.'' First, there was a break in the continuity of leadership, with the initial team of founding editors becoming significantly less involved between 2005 and 2007 until a second generation of admins had taken over. This second generation of admins is the group ultimately accused of ``capturing'' the Croatian Wikipedia and had decidedly different priorities. P06, a Croatian Wikipedia editor, described the leader of this second generation, Admin 1,\footnote{While the usernames of these admins are publicly available, they have been anonymized for this paper.} as someone who wanted to ``rule Wikipedia.'' P10 further noted that Admin 1's primary goal was not to create an encyclopedic project, but rather to foster an ``inner group cohesion'' among a core group of contributors. This focus on inner group cohesion directly led to the formation of an insular bureaucratic culture on Croatian Wikipedia, where the only avenue through which contributors could ascend to leadership positions on the project was through personal relationships with this core group of admins. As a result, and in contrast to the open bureaucratic culture of early Serbian Wikipedia, P10 noted that for a period on Croatian Wikipedia, Admin 1 was ``practically the only active bureaucrat'' on the project.

\subsubsection{The Rise of a Cabal on Croatian Wikipedia.}
\label{cabals}
The early insular bureaucratic culture of Croatian Wikipedia was further codified into closure through the formation of what Wikipedians refer to as an administrator ``cabal'': a select group of users, often ones with elevated privileges, that coordinate off-project with the goal of synchronizing their on-project positions in a non-transparent manner. P12, a steward with experience in multiple conduct enforcement cases on Wikipedia, explained that the primary cabal on Croatian Wikipedia formed as Admin 2, the second prominent member of the second generation admin team, ``quickly developed what is best described as a personality cult.'' Later on, this user established off-wiki communication channels on various platforms, particularly Discord and Telegram. 

Checkusers on Wikipedia have access to technical data stored by the server about user accounts, including IP addresses, and are thus a project's first line of defense for detecting and combating coordinated and abusive behavior.\footnote{\url{https://meta.wikimedia.org/wiki/CheckUser_policy} (\url{https://perma.cc/BZF5-CT22})} P12 explained that through these off-wiki channels, Admin 2 ``installed a bunch of friendly admins. He also installed two checkusers, one of whom would eventually fabricate checkuser data to try and secure his innocence.'' By limiting bureaucrat and checkuser positions to personally favored contacts, the members of the cabal eliminated a critical check on their power. This allowed them to engage in behavior that violated both the content and conduct principles of Wikipedia without reproach. 

P09, a former Croatian Wikipedia editor, further described how the formation of this cabal negatively affected communication among contributors on the project: ``You have a really small group of people who have control, who communicate internally without transparency, who synchronize their positions, and only then present them as a kind of univocal, more or less, solution, although you know, that this is not the state of affairs.'' Rather than unfolding on-wiki, where all members of the project could participate in deliberation, Croatian Wikipedia's decision-making process was relegated to private chats between a select group of users.

Cabals themselves are not uncommon on Wikipedia, as P12 pointed out.\footnote{Accusations of cabal rule are so commonplace in editor disputes on talk pages that a satirical page exists with an extensive list of evil cabals that supposedly control the online encyclopedia: \url{https://en.wikipedia.org/wiki/Wikipedia:List_of_cabals} (\url{https://perma.cc/HY2T-A2L7})} Though transparent decision-making and public deliberation is valued, there are cases where, private, off-wiki communication among users is preferable, such as in collecting evidence prior to publicly bringing up an arbitration case. But P12 noted that key decisions that affect the project should never be made in private channels. Furthermore, while user cabals exist across projects, P12 assessed that they pose little risk on more mature projects, such as English Wikipedia, because these projects already have ``existing power structures in place'' that are difficult to subvert by cabal rule. Croatian Wikipedia's capture seems to have been set in motion before such a structure was fully established.

\subsubsection{Community Capacity to Respond to Governance Threats.}
\label{diverging-capacity}

The establishment of cabal rule on Croatian Wikipedia significantly raised barriers to contribution and limited opportunities for contributors to ascend to leadership positions. In the words of several interviewees, the project was ``locked'' to contributors outside of the cabal. These barriers to contribution depleted the community's capacity to respond to abuse on the project. 

Interviewees in global and cross-wiki roles stressed how critical community capacity was to combating disinformation and related threats. ``If you have a small number of editors, you only have a certain number of people who can deal with vandalism and all of the rest of it,'' P11, a WMF staff member who works on trust \& safety issues, explained. P04, a Wikipedian who led a cross-wiki volunteer project trying to root out climate change denial across language editions, noted that Wikipedia generally has ``really good mechanisms to deal with disinformation. But all of those mechanisms require editors, and a lot of them. You need to make sure that there's always somebody that can check that you're adding the correct information.'' P04 noted that smaller Wikipedias, as well as the less visible articles on English Wikipedia, are less likely to have the ``critical mass of editors'' needed to constantly monitor for disinformation and conduct issues.

Volunteers fighting abuse on their projects are at a fundamental disadvantage: protecting a project from vandalism, trolling, and disinformation campaigns can be significantly more resource-intensive than attacking a project. While a variety of technical tools exist on Wikipedia to allow volunteers to quickly detect and revert problematic or abusive edits, these tools are insufficient for dealing with more subtle and sophisticated attacks. This imbalance is exacerbated by the fact that relatively few volunteer contributors have the time, interest, or expertise necessary to perform these functions, often preferring to contribute content to the project in the form of articles rather than policing and vetting new contributions. Because of these challenges, community capacity is an important factor in how well a project can respond to threats to both its content and governance mechanisms.

It is worth noting that Serbian Wikipedia has a naturally larger editorial base than Croatian Wikipedia in part because its language community is larger. Roughly 7.3 million people speak Serbian as their mother tongue, compared to the 4.9 million who speak Croatian \cite{worlddata.info_2023, worlddata.info_2023a}. Nonetheless, Croatian's bureaucratic closure exacerbated the disparity, with several interviewees noting that a sizeable portion of editors simply stopped being active on Croatian Wikipedia because of the hostile environment. Furthermore, the current administrators of the Croatian Wikipedia, who were elected following the ``unlocking'' of the project, report that the scale of the damage done to the project under the rule of the cabal continues to hamper its development. P05, a newer admin on the project, noted that they spend a significant amount of time ``trying to keep Croatian Wikipedia unvandalized,'' which detracts from their ability to contribute to other areas the project sorely needs, such as rule development.

In addition to severely limiting the capacity of the community to deal with disinformation-related threats, Croatian's bureaucratic closure sapped the project of the opportunity to develop another critical resource: a diversity of views. P08 explained:

\begin{quote}
Croatian Wikipedia, sort of made a cabal: it made an ideologically similar clique of people that sort of self perpetuated themselves. And here, on Serbian Wikipedia, we just didn't do that. It was never one clan or the other that had an advantage or upper hand. We were always a mixed bunch. And I think that helped with intervening when things got messy. And that's why they never escalated so much.
\end{quote}

\noindent P07 echoed this observation about the ideological composition of Serbian Wikipedia, stating, ``there was a strong polarization between liberals and right wingers. It was really like 50-50.'' 

The reference to ideological polarization among the editorial base of Serbian Wikipedia as a positive feature may appear somewhat counter-intuitive, as editors with opposing political views are presumably more likely to disagree with one another, leading to a more conflict-prone environment on the project and potential ``edit warring'' \cite{sumi.etal_2011}. But in fact, the observation about the positive role of polarization in the Serbian Wikipedia community echoes the findings of \citet{shi.etal_2019} in ``The Wisdom of Polarized Crowds:'' politically polarized teams of editors create better quality articles than politically homogeneous teams on topics related to politics, social issues, and science. The proposed reasoning for why this is the case is that polarized groups of editors have longer, more argumentative debates in talk pages, and this process of argumentation ultimately leads to more robust articles that consider and present multiple points of view rather than privileging one partisan viewpoint over others.

\subsection{Proposition 3: Formal Institutional Organization}
\label{proposition-3-formal-institutional-organization}

A final theme that participants cited as contributing to governance capture concerned the degree to which the project developed more formal forms of organization that provided mechanisms for internal accountability from within the project and external scrutiny from the broader Wikimedia movement. Two types of formal organization were cited as particularly important to a project's resilience to capture: formal rules, particularly rules constraining the power of administrators, and the establishment of relationships with external community groups and partnerships related to the Wikimedia movement.

\subsubsection{Rules Constraining Admin Power.}

Interviewees noted either the underdevelopment, in the former case, or the absence, in the latter, of two rule pages on Croatian Wikipedia that are present on English Wikipedia and other large projects with mature rules environments: WP:Admin and WP:Block. The first of these polices, WP:Admin, typically outlines the role and responsibilities of administrators on a project, and, critically, checks on an administrator's power. For example, on the English Wikipedia, the WP:Admin policy specifies the processes for reversal of an administrator's action and removal of adminship in cases where an admin abuses their privileges.\footnote{\url{https://en.wikipedia.org/wiki/Wikipedia:Administrators} (\url{https://perma.cc/6UVS-WDRM})}

While each project establishes policies such as WP:Admin autonomously, most medium-to-large projects adopt some type of policy to govern the conduct of administrators and to institute local mechanisms to remove an administrator who violates these policies. P01, a steward, described how this process works on their primary Wikipedia edition, a medium-sized project:

\begin{quote}
There's an established system of both appointing people to administrator roles on a project, and also an established system of removing people from the role, which essentially means that if, if a community member feels that I am no longer a good fit as [my primary language edition] Wikipedia administrator, they can start a process that at the end requires me to stand for reelection and prove that I still have the trust of the community. This usually is not the case on projects that exists for a smaller amount of time than [my primary language edition] Wikipedia and haven't have the chance to develop those procedures yet, or projects that are simply smaller, and only consist of two, three people who know each other very well and, don't even feel the need for such policies.''
\end{quote}

The WP:Admin page on Croatian Wikipedia, while present, had no such processes specified as of the June 16, 2021 version of the page. Instead, the page included a few paragraphs on the general role of admins on Wikipedia; a quote from Wikipedia founder Jimmy Wales discouraging treating admins as all-powerful authority figures; and a list of all former and current admins on Croatian Wikipedia.

\begin{figure}[t]
    \centering
    \includegraphics[width=0.75\textwidth]{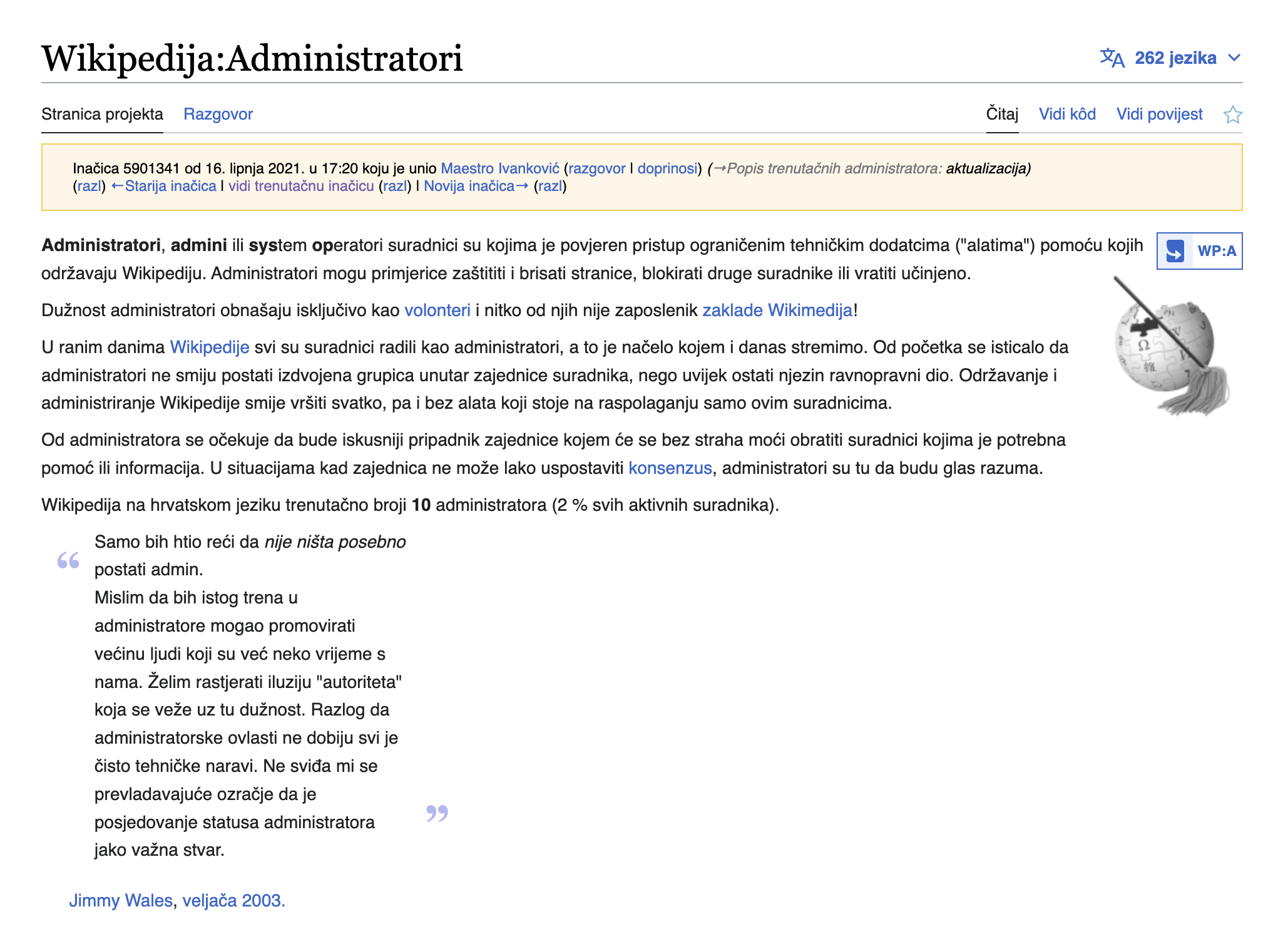}
    \caption{Screenshot of Croatian Wikipedia WP:Admin page as of June 16, 2021, showing a general description of the role of administrators but no guidelines on administrator conduct, as is found on the English Wikipedia.}
    \label{fig:hrwiki}
\end{figure}

In addition to outlining the expected roles of administrators and tools available to them, Serbian Wikipedia's WP:Admin page outlinks to an additional policy on the voting process and rules for granting as well as revoking administrator status in case where the admin or bureaucrat abuses their rights or the community loses confidence in their ability to carry them out.\footnote{\url{https://sr.wikipedia.org/sr-el/Википедија:Захтеви_за_администрирање/Правила_гласања} (\url{https://perma.cc/XL6C-WUWY})}
Importantly, this page also imposes minimum conditions on voters in admin elections, such as having an active account for at least three months and having made a certain number of recent edits. These conditions limit the ability of users to rig admin elections through the use of sockpuppet accounts, a tactic that was commonly employed by the admin cabal on Croatian Wikipedia.

Following the ousting of the right-wing admin cabal, Croatian Wikipedia did begin the process of developing these policies. The most recent diff of the WP:Admin page at the time of analysis, as of July 19, 2022, does list basic policies governing the conduct of admins.\footnote{\url{https://hr.wikipedia.org/w/index.php?title=Wikipedija:Administratori&oldid=6429986} (\url{https://perma.cc/NV6C-QY7F})} Although the page in its most recent form is not as developed as that of English Wikipedia and other large Wikipedias, the addition of these basic rules reflects improvements to the Croatian Wikipedia rules environment by the current admins and active contributors to the project following the removal of the right-wing admin cabal. These improvements were corroborated in our interviews. P13, for example, noted that the project's ``atmosphere is definitely different. You can actually work now on Croatian Wikipedia.'' He described how his work on an article about minority languages in Croatia was welcomed by the new admin cohort, whereas it would be ``impossible'' to write such an article under the previous admins. He partly attributed the difference to the more mature rules environment, noting that, previously ``it wasn't that bureaucratized, it was just, you know, that there would be, you have a couple of influential editors to whom the community gave the trust, let's say, and they were free to do more or less anything to block you, to disrupt you, to do whatever they want.''

The second policy, WP:Block, describes how admins are to implement blocking: the act of technically preventing user accounts or certain IP addresses from editing Wikipedia. On English Wikipedia, where this policy is thoroughly documented, blocking is described as a tool to prevent disruption or damage to Wikipedia, ``not to punish users.''\footnote{\url{https://en.wikipedia.org/w/index.php?title=Wikipedia:Blocking_policy&oldid=1098590974} (\url{https://perma.cc/7EC2-DVTU})} Serbian Wikipedia's WP:Block outlines a similarly extensive policy with the same clarifications on the role of blocking and when administrators can use it.\footnote{
\url{https://sr.wikipedia.org/w/index.php?title=Википедија:Правила_блокирања&oldid=24135057} (\url{https://perma.cc/N5XJ-HPPN})}

Croatian Wikipedia has also historically lacked a documented blocking policy. Interviewees noted that blocks were used liberally by admins against users that critiqued their actions or made edits that did cohere with the ideological leanings of the admin cohort. As of October 31, 2022, Croatian Wikipedia had a project page that outlined a proposal for WP:Block, but the proposal was listed as still under discussion and had not yet been implemented by the community.\footnote{\url{https://hr.wikipedia.org/w/index.php?title=Wikipedija:Pravila_blokiranja&oldid=6196463} (\url{https://perma.cc/QQ6J-Z29Y})} In the absence of documented policies constraining admin behavior, members of the admin cabal invoked rules and policies to silence those on the project that criticized their activities. P14 described this environment aptly:

\begin{quote}
They went through great lengths to organize a system where you're so easily proclaimed a bot, you're proclaimed somebody who's abusing the, you know, whatever\ldots{} they found these excuses to institutionalize their order of how things are done.
\end{quote}

Other interviewees, like P06 and P09, recalled specific examples of members of the admin cabal invoking rules in a manner that had chilling effects on good-faith contributors. P06 recalled an instance during which an admin blocked an editor for pointing out the far right bias of Croatian Wikipedia, citing a policy against hate speech. P09, meanwhile, noted that problematic admins on Croatian Wikipedia ``tended to use very technical ways to kind of suppress new users and content,'' describing how their edits to articles on ``progressive issues,'' particularly LGBT-related content, were monitored by admins through the recent changes log and repeatedly deleted based on language or style-related technicalities.

One editor, P06, explicitly likened Croatian Wikipedia's model of governance to feudalism:

\begin{quote}
We have admins who behave like medieval lords, and we are basically serfs. And they're not bound by the rules, they are doing everything by themselves, how they deem necessary. And the only thing they fear is that another admin will step in and maybe argue with them about their behavior towards other users. That's the only thing, because there are no rules.
\end{quote}

In the absence of institutional constraints on administrator behavior in the form of formal rules, Croatian Wikipedia defaulted to a reliance on personal relationships and deference to the absolute power of administrators as a means of making policy decisions, a community management strategy that echoes the ``implicit feudalism'' proposed by \citet{schneider_2022} as well as what \citet{frantz2021personalist} describes as ``personalistic'' governance styles associated with democratic backsliding in traditional institutions. In contrast, P07 said the Serbian Wikipedia possesses the ``densest due process.''

\subsubsection{The Presence of Well-Organized Community Groups and Partnerships.}
In addition to the rules environment, interviewees pointed out another area where Croatian Wikipedia's institutional maturity lagged behind other language editions: the project's integration with other elements of the Wikimedia movement. While the core of Wikipedia is its volunteer editing community, the broader Wikimedia movement is much more expansive, encompassing an array of activities, projects, and organizations.\footnote{\url{https://meta.wikimedia.org/wiki/Wikimedia_movement} (\url{https://perma.cc/HP8E-C5DL})} As a result, there are many ways to contribute to the Wikimedia movement other than editing articles.

P09, who has a background in community organizing, stated that one of the fundamental differences between the trajectory of Croatian Wikipedia and that of Serbian Wikipedia is that the latter had a more robust community interested in education and outreach, rather than solely on editing articles:

\begin{quote}
Unlike the Serbian community, which actually kept progressing in terms of resources, kept developing in terms of different forms of work\ldots{} Croatian Wikipedia never did projects, they never developed partnerships. They never developed outreach methods, educational methods, any of these things. So that was a huge problem. Basically you had only people who will do one type of work within Croatian Wikipedia, and that's editing pages. You will not have people who were passionate about photography, who were passionate about style, who were passionate about outreach and educating others.
\end{quote}

\noindent P09 considers these efforts essential to attracting newcomers from backgrounds in civil society who have experience in community organizing and management, as well as those from demographics who are underrepresented on Croatian Wikipedia, such as women and LGBT editors.

In addition to these efforts to engage in different forms of movement-building, interviewees also mentioned the role of more formal organizational structures, such as local Wikimedia chapters. P13 and others noted that the presence of a well organized local chapter, Wikimedia Serbia, was one of the primary reasons why Serbian Wikipedia has more developed online community governance than the other four projects in BCMS environment. As P01, P12, and several other interviewees noted, these chapters do not tend to become involved in content governance on their associated language editions. But because they are legal non-profit entities, the process of incorporating and maintaining an active Wikimedia chapter invites scrutiny into the project that encourages the development and enforcement of community policies in line with Wikipedia's principles. The additional visibility afforded by the presence of a chapter that engages in outreach and public relations also makes it less likely that attempts at project takeover would go unnoticed by functionaries outside of the project.

\section{Discussion}
\label{discussion}

The primary contribution of this paper is to show how different self-governed communities---built on the same underlying software, situated in a common sociolinguistic environment, and constituting what appear to be similarly attractive targets for influence actors seeking to shape a broader audience's perception of historical events---diverge in outcomes because of key differences in organizational and institutional design. In the subsections that follow, we introduce a conceptual model to synthesize our findings and discuss its applicability to Wikipedia projects beyond the BCMS editions. We also contrast our findings with related work discussed in §\ref{sec:related-work} about factors that facilitate growth and success in online communities more broadly, and consider the implications for self-governed online communities beyond Wikipedia.

\subsection{A Conceptual Model of Institutional Capture Risk}

\begin{figure}[t]
    \centering
    \includegraphics[width=1.0\textwidth]{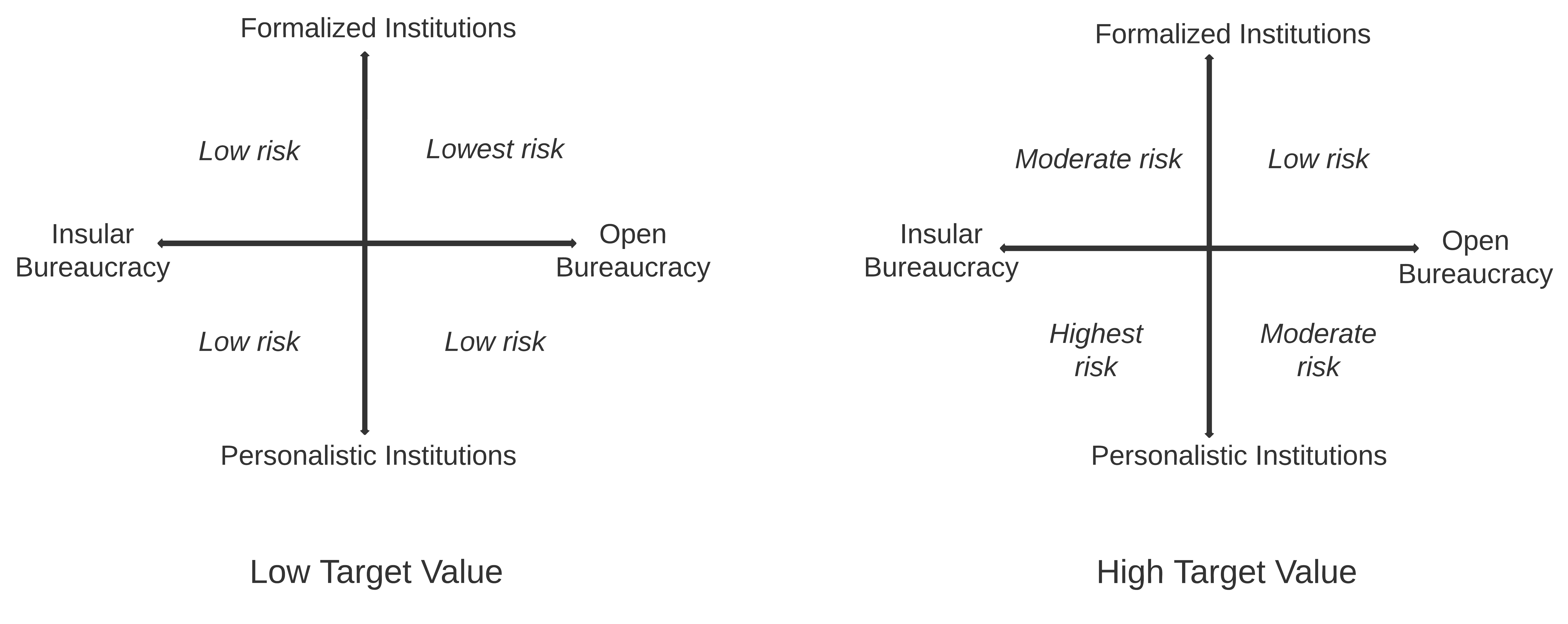}
    \caption{A conceptual model that visualizes possible institutional configurations for Wikipedia projects that affect the risk of governance capture, derived from our three propositions. The bottom left quadrant of the second 2x2 matrix, representing projects with high value as a target, insular bureaucracies, and personalistic institutions, provides the largest ``window of opportunity'' for governance capture.}
    \label{fig:capture}
\end{figure} 

Our results imply a conceptual model in which each of our three propositions describe positions that increase the risk of governance capture. We present a graphical display of this conceptual model in Figure \ref{fig:capture}. Proposition 1 (§\ref{proposition-1-percieved value}) is visualized as the difference between the sets of quadrants on the left and right and shows how high perceived value as a target---largely in the form of large audiences and nationalistic resonance in our context---increases risk of capture. If a project is not seen as valuable to control by potential information operations, actors are simply unlikely to want to capture it. This proposition in isolation might be able to explain why the Serbo-Croatian and Bosnian Wikipedias were not captured. Since both Serbian and Croatian were described by our interviewees as similarly valuable, however, it cannot explain the relatively different outcomes in these two projects.

The $y$ and $x$ axes of the two subgraphs represent Proposition 2 (§\ref{proposition-2-bureaucratic-openess}) and Proposition 3 (§\ref{proposition-3-formal-institutional-organization}), respectively. As a result, each of the four quadrants in the matrix represents one of four possible configurations of a project along these two dimensions: 

\begin{itemize}
\item Top left: Insular Bureaucracy but Formalized Institutions; 
\item Top right: Open Bureaucracy and Formalized Institutions; 
\item Bottom left: Insular Bureaucracy and Personalistic Institutions; and
\item Top right: Open Bureaucracy but Personalistic Institutions.
\end{itemize}

Our results in §\ref{proposition-2-bureaucratic-openess} suggest that Serbian Wikipedia invested early in bureaucratic openness. Given the risks that greater openness brings with it---exposing the project to an increasing number of damaging contributions alongside valuable contributions---it may seem counter-intuitive that early bureaucratic openness served as a protective measure. But Serbian Wikipedia coupled its investment in openness with an investment in formal forms of organization both on and off the project. On the project, Serbian Wikipedia developed rules that governed, and crucially, circumscribed, administrator behavior. The project also pursued off-wiki avenues of formal organization, such as by establishing a local Wikimedia chapter. These additional forms of organization provided both greater accountability on the project and directed a degree of internal and external scrutiny to it.

Croatian Wikipedia followed a different path. Rather than investing in bureaucratic openness and institutional formalization, it appears to have neglected both in its early stages. Instead, it opted for an over-reliance on informal norms and personal relationships as organizing principles. As the project grew and increased in perceived value, a faction of editors, driven by a mix of ideological and interpersonal motives, were presented with a window of opportunity to consolidate their control over the project. The institutional context of Croatian Wikipedia provided little indication that there would be significant costs imposed on them, either locally or through Wikipedia's global governance mechanisms. The editors seized this opportunity, and in doing so, established a decade-long rule over the project.

If we look only within the rightmost subdiagram of Figure \ref{fig:capture} where both Croatian and Serbian Wikipadias would be located, our findings suggest that projects in the bottom left quadrant (such as Croatian Wikipedia) will be at the highest risk. Projects in the top right quadrant (such as Serbian Wikipedia) will have low risk. Projects in the remaining two quadrants will each have medium levels of risk. Because a project with low perceived value is likely undesirable in itself, the most desirable configuration for projects, in terms of resilience to project capture, is thus to have a relatively open bureaucratic structure but with formalized institutions. 

\subsection{Implications for Knowledge Integrity Work}

Researchers studying online disinformation have tended to focus on identifying features of problematic content and arresting the processes through which it is disseminated. Accordingly, many of the solutions currently under development to address disinformation campaigns and other influence operations in the context of Wikipedia involve the introduction of automated tools to detect problematic content and behavior \cite{lucassen.schraagen_2011, sathe.etal_2020, potthast.etal_2008, mola-velasco_2011}. These tools may empower good faith administrators to fight ``one-off'' risks like vandalism more efficiently, but they do not address the more fundamental question of how various institutional arrangements condition how power is configured in online communities, and how those dynamics in turn affect information integrity outcomes. 

While Croatian Wikipedia remains the most well-known case of governance capture, a number of Request for Comment (RFC) processes over the last decade have documented similar struggles on other Wikipedia language editions. For example, in a now-closed RFC titled, ``Do something about azwiki,''\footnote{\url{https://meta.wikimedia.org/wiki/Requests_for_comment/Do_something_about_azwiki} (\url{https://perma.cc/Y2GU-R4TV})} editors document a series of issues with the local governance of Azerbaijani Wikipedia that bear a striking similarity to those of Croatian Wikipedia: the abuse of administrative block tools, sockpuppetry, and the regulation of key decision-making about community governance to private, off-wiki channels, such as WhatsApp and Facebook Groups, steered by a few individuals. Similar to the Croatian Wikipedia, the RFC also connects this behavior to a pattern of systematic historical revisionism on the project, particularly in articles about the Armenian Genocide. While the RFC was closed in July 2019 and resulted in the removal of one problematic administrator, the last update posted by a steward on the RFC acknowledged that the problems of governance on the Azerbaijani Wikipedia would require deeper engagement to address. In addition to Azerbaijani Wikipedia, governance arrangements similar to those on Croatian Wikipedia have been publicly reported in varying levels of detail on the Japanese Wikipedia \cite{sato_21, kim.etal_2023}, Chechen Wikipedia\footnote{\url{https://meta.wikimedia.org/wiki/Requests_for_comment/Massive_sysop_abuse_in_Chechen_Wikipedia} (\url{https://perma.cc/A2S5-DGVL})} and Hindi Wikipedia.\footnote{\url{https://meta.wikimedia.org/wiki/Requests_for_comment/Userrights_on_Hindi_Wikipedia} (\url{https://perma.cc/64L6-6PNN})}

\subsection{Implications for the Study of Self-Governed Communities}

Studies of governance and moderation on Wikipedia and self-governed communities more broadly have identified technical and social configurations that can foster growth and success in these communities. These include the establishment of rewards and sanctions and lowering entry barriers for newcomers. Although these factors appear similar to the propositions of bureaucratic openness and institutional formalization presented in this paper, they are not helpful in the case of Croatian Wikipedia. For example, in their chapter on ``Regulating Behavior in Online Communities,'' \citet{kiesler2012regulating} discuss how outside ``manipulators'' may violate a community's behavioral norms by artificially boosting a business's rating on Yelp or editing a Wikipedia article to reflect their point of view. They do not consider, however, how insider members may exploit vulnerabilities in a community's self-governance systems to define \textit{how} rules are shaped and \textit{who} gets to shape them. In distinguishing state capture from corruption and other forms of political influence, \citet{hellman.etal_2003} note that the former is the shaping of the ``basic rules of the game'' for private gain, as opposed to merely arranging for existing state policies to be applied in a favorable manner to the elites in question. 

Platforms that enable elements of self-governance in their institutional design afford greater autonomy to communities to design governance structures according to their needs \cite{schneider.etal_2021}. But they also open the door for those very structures to be exploited by motivated actors. Thus far, governance capture remains almost entirely unstudied in the social computing literature, even though the governing structures of online communities, like those of their offline counterparts, remain inherently vulnerable \cite{schneider_2022}. On Reddit, where individual subreddits establish their own community guidelines enforced by local administrators, \citet{massanari_2017} describes the rise of ''mini-fiefdoms''---clusters of related subreddits promoting ``toxic technocultures'' such as GamerGate---controlled by a very few number of moderators. Mastodon's model, which consists of individual servers each run by an administrator that sets its own policies, \cite{gehl.zulli_2022}, also allows for the emergence of even more decentralized forms of self-governance. Amid increasing interest in decentralized social media as legitimate alternatives to mainstream commercial platforms like Facebook and Twitter \cite{perez_2022}, more attention should be paid to the vulnerabilities inherent in these models of self-governance.

It is not our goal to design these vulnerabilities \textit{out} from sociotechnical systems. In fact, in ``Anarchy, Status Updates, and Utopia,'' \citet{grimmelmann_2014} argues that `Social software has a power problem...There is no way to redesign the technologies of social software so that technical power disappears, for the reason that it is the social power that gives the technical power its bite.'' But as this study has shown, variation in self-governing structures at the community level can indeed result in divergent outcomes for otherwise similar communities. This suggests that, at the very least, certain governance arrangements can provide checks on the accumulation of both technical and social power in online institutions, ultimately resulting in more democratic, participatory, and resilient self-governed communities. Our work also suggests that this can reduce other social problems like disinformation. 

\section{Limitations}
\label{sec:limitations}

An important limitation of our study is that none of its authors are fluent BCMS speakers. As a result, interviews were conducted in English. While English proficiency is very high in Croatia and high in Serbia as well \cite{efeducationfirst_2022}, this reduced our pool of potential interviewees and meant that certain perspectives were likely not represented in our interview data. In particular, there is likely a subset of contributors on the Croatian and Serbian Wikipedias that do not regularly engage in global governance discussions, which tend to occur in English, but that still have insight into the local dynamics of their respective projects. 

In an attempt to partially address the language limitation, we machine translated (via Google Translate), read, and coded talk pages from the Serbian and Croatian Wikipedias where contributors were specifically discussing governance on the projects. While we acknowledge machine translation is flawed, we used it as a method of triangulation for our rich interview dataset rather than as a source of primary information. The themes that surfaced in these talk pages matched themes that emerged in our interview data, increasing confidence in our findings. 

Another related set of limitations stems from missing perspectives in our data. Although we reached out to a broad array of individuals, not everyone we reached out to responded to our interview request. In particular, we contacted two former Croatian Wikipedia administrators who were said to have played key roles in the capture of the project but did not receive a response from either. 

While the individuals we were able to interview had a range of backgrounds and experiences on the platform, they also shared a few characteristics. First, they tended to be more involved in community governance processes than the average Wikipedia contributor and were eager to talk about these processes. Second, they generally tended to view disinformation and governance capture as threats to Wikipedia; in many cases, they devoted a significant amount of their volunteer time and resources to combating these threats. This might not be a widely shared perspective within the BCMS Wikipedias or among Wikipedians more broadly.

A third related limitation is that while our findings describe processes that play out over many years, this broader perspective relies on triangulation across interviewees. For example, while descriptions of differences in bureaucratic culture between Serbian and Croatian Wikipedia were nearly universal, the specific aspects individuals described varied based on the time period and nature of their engagement on either project. For example, P08 and P10 (users who assumed leadership positions in the Serbian Wikipedia within two years of the project's founding) and P06 and P14 (who began editing Croatian Wikipedia soon after its founding) provided insight into how the early bureaucratic cultures of the two projects differed. Other interviewees with experience later on in the projects' lifecycles (like P09, P11, and P15) corroborated the early participants' assessments about the differences in bureaucratic culture and provided additional insight on the effects on the respective communities' capacities to deal with threats that strained existing governance processes (§\ref{diverging-capacity}). 

Our work is also limited in that our selection of cases held things constant which might also matter. For example, assessing Wikipedia language editions according to our three propositions can allow for cross-comparison of relative risk in a manner that is language agnostic, but does not preclude consideration of sociocultural context in ways that \citet{khatri.etal_2022} have shown are important drivers of activity. 

It might also be that Croatian Wikipedia reflects only one potential path. For example, the WMF's Trust \& Safety team put out a statement in September 2021 about ``infiltration concerns'' on the Chinese-language Wikipedia and announced a decision to globally ban a number of administrators as the result of an internal investigation. Subsequent reporting alleged that the administrators in question promoted a ``pro-Beijing viewpoint,'' meddled in administrator elections, and engaged in abuse and harassment of other volunteers on the project that escalated, in one case, to physical assault \citep{pasternack_2021}. Of course, the People's Republic of China has instituted restrictions on its citizens' access to Wikipedia since 2004, blocking the Chinese-language edition in 2015 and the entire site in 2019 \citep{harrison_2019}. The addition of government pressure within the project would likely influence how capture manifested. Future research could extend our framework to better reflect the role of external political factors that were absent from the Croatian case.

Lastly, like all grounded theory-based work, we are limited in our ability to generalize to other contexts. Although we believe that insights from our study will be useful for understanding governance capture dynamics in other self-governed online communities, our approach sacrifices breadth for depth in data collection in a way that means that knowing for certain whether similar dynamics will play out in other contexts must be left for future work. We hope others conduct research on these dynamics using a range of methods---including quantitative hypothesis testing approaches---and in a range of other settings.

\section{Conclusion}
\label{conclusion}

As early 2000s enthusiasm about the internet has largely given way to cynicism and concern, Wikipedia's largely warranted reputation as the ``last best place on the internet'' has served as an important counterpoint \cite{cooke_2020}. The capture of Croatian Wikipedia and its reconfiguration as a source of nationalist disinformation provides reason to believe that Wikipedia's success may be fragile.
But in that Croatian Wikipedia reflects a relative outlier among other BCMS Wikipedias with so much else in common, it also reflects an opportunity to learn through systematic comparison. Our work offers one set of explanations for Croatian Wikipedia's capture. In doing so, we demonstrate how comparative studies and a focus on project governance provide opportunities for new insights into the widely studied problems related to knowledge integrity and disinformation. We hope that our approach can serve as a productive and useful tool in the fight against disinformation going forward.


\begin{acks}

Earlier versions of this paper benefited from valuable feedback from Pablo Aragón, Diego Sáez Trumper, Jonathan Morgan, and members of the UW Community Data Science Collective. In addition, thank you to the members of the student research team that helped with the initial transcription and coding of the interviews: Ashlyn Aske, Richard Lei, Chris Lyu, and Yimeng Wang. We also thank the anonymous reviewers and program committee members at CSCW for their helpful guidance in improving this paper. Finally, we owe special gratitude to the 15 interview participants who shared their time, experiences, and knowledge with us. Financial support for this work came from the US National Science Foundation (awards IIS-2045055 and IIS-1749815).

\end{acks}

\bibliographystyle{ACM-Reference-Format}
\bibliography{wiki_references}


\appendix{}

\section{Interview Protocol}
\label{sec:protocol}

\subsection{Background}

\begin{enumerate}
\item Can you tell me about your personal background and history of involvement with Wikipedia?
\item What was your experience with [PROJECT NAME]? When did you get involved, and in what capacities? 
\item How long were you active on the project?
\end{enumerate}

\subsection{General perceptions of disinformation on Wikipedia}
\begin{enumerate}
    \item Do you think disinformation is a problem or a threat for Wikipedia? Why do you feel that way? 
    \item  How would you compare Wikipedia’s efforts to counter disinformation with that of big commercial social media platforms? 
    \item On the Wikipedia projects you are active in, have you witnessed or heard of any attempts to introduce systematic content bias to articles? 
    \item Have you encountered any cases of historical revisionism or propaganda on language projects you are active in? Can you provide some examples?
    \item What sorts of tactics do rogue administrators use to insert content bias?  
    \item How effective is the Meta RfC process in resolving content bias disputes on Wikipedia language editions? 
\end{enumerate}

\subsection{Pertaining to a single language edition}
\begin{enumerate}
    \item During the period you were active on [PROJECT A], how would you describe the environment?
    \item What was your relationship like with the administrators of the project at that time? How free did you feel to make contributions to the project?
    \item What policies does [Project A] have in place to protect against content bias?
    \item How were these policies enacted in practice?
    \item Could you describe situations where [Project A] had issues with content bias in articles, and how it handled them?
    \item How does [Project A] handle content bias in articles about regional culture and history, specifically? 
    \item How does [Project A] handle content bias in articles about socially controversial issues, specifically? 
    \item How ideologically diverse is [Project A’s] community of active contributors? 
    \item How ideologically diverse is [Project A’s] moderation team? 
    \item What kind of automated fact-checking/vandalism detection tools does [Project A] use? How effective are they?
    \item Overall, how would you describe the quality of articles within [Project A]?
    \item How does [Project A] deal with cases of rogue administrators? 
    \item How does [Project A] deal with cases of off-platform harassment of users with dissenting views, or other abusive behavior?
\end{enumerate}

\subsection{Cross-project comparisons}
\begin{enumerate}
    \item Can you tell me about projects that you were involved in that faced systemic or repeated content bias issues? 
    \item Can you tell me about projects that you were involved with that faced issues with rogue administrators?
    \item If you are active in or monitoring multiple Wikipedia language editions, which ones are better at preserving a Neutral Point of View (NPOV)? Why do you think that is?
    \item What level features do you think make certain projects more vulnerable than others to abuse?
\end{enumerate}

\subsection{Additional questions for Wikimedia researchers and stewards}
\begin{enumerate}
\item How do you think about content-based threats (affecting articles) vs. behavior-based threats (affecting interactions among editors, project rules, etc.)? Is there a distinction, if any?  
\item When do you “step in” to support a language edition that is dealing with disinformation issues or potential capture? How did this process work? What difficulties do you tend to encounter? 
\item Can you tell me about cases where you have had to deal with “project infiltration” by an external actor (someone who is not an editor of the project)? 
\end{enumerate}

\end{document}